\documentclass[%
aip,
 amsmath,amssymb,
 reprint,%
]{revtex4-1}
\usepackage{accents}
\usepackage{xcolor}
\usepackage{subfigure}
\usepackage{graphicx}% Include figure files
\usepackage{dcolumn}% Align table columns on decimal point
\usepackage{bm}% bold math
\usepackage{appendix}
\usepackage{color}
\begin{document}

%\preprint{APS/123-QED}

\title{Different responses of the Rayleigh-Taylor type and resistive drift wave instabilities to the velocity shear}% Force line breaks with \\
%\thanks{A footnote to the article title}%

\author{Y. Zhang}
 %\email{yaz148@ucsd.edu}
\affiliation{%
 Mechanical and Aerospace Engineering Department, University of California San Diego, La Jolla, CA 92093, USA
}%\\
\author{S. I. Krasheninnikov}%
 %\email{skrash@mae.ucsd.edu}
\affiliation{%
 Mechanical and Aerospace Engineering Department, University of California San Diego, La Jolla, CA 92093, USA
}%

\author{A. I. Smolyakov}%
 %\email{skrash@mae.ucsd.edu}
\affiliation{%
University of Saskatchewan, Saskatchewan, Saskatoon SK S7N 5E2, Canada 
}%
%\date{\today}% It is always \today, today,
             %  but any date may be explicitly specified

\begin{abstract}
The effects of velocity shear on the unstable modes driven by the effective gravity  (Rayleigh-Taylor and interchange) and resistive drift wave instabilities for inhomogeneous equilibrium fluid/plasma density are analyzed for the localized eigenmode problems. It is shown that the effect of the velocity shear drastically depends on the type of instability. Whereas the velocity shear can significantly suppress both Rayleigh-Taylor and interchange instabilities, it has only a week impact on the growth rate of the resistive drift wave. This is directly related to the physical nature of these instabilities. For the Rayleigh-Taylor and interchange instabilities, the shear flow tilts the eddies of the stream functions, while for the resistive drift wave instability the shear flow simply shifts the eddies in the radial direction with no tilting. However, for large velocity shear, the eigenmode solutions for resistive drift waves cease to exist.
\end{abstract}

\maketitle
There is a long-lasting interest in the impact of the velocity shear on fluid and plasma instabilities and turbulence (e.g. see Refs. \onlinecite{kelvin1887stability, kuo1963perturbations,chandrasekhar2013hydrodynamic, hassam1992nonlinear,benilov2002does,lehnert1966short,sugama1991radial,miller1995stabilization,carreras1993resistive,diamond2005zonal,waelbroeck1992theory}). In fusion research this interest is stipulated by numerous experimental evidence of the role of strong shear of plasma flow on the transition of plasma operation from a low confinement L-mode to a high confinement H-mode\cite{ritz1990evidence,groebner1990role,burrell1997effects,van2003turbulent,fujisawa2011experimental}. Underlying physics of these phenomena are poorly understood and, therefore, further theoretical studies are necessary,  e.g., see Refs. \onlinecite{carreras1993resistive,sugama1991radial,miller1995stabilization}). 

It is often assumed (e.g. see Refs. \onlinecite{horton2012turbulent,kinsey2005nonlinear,waltz1997gyro}) that the plasma instabilities and fluctuation levels will be quenched if the shear of plasma flow velocity, $|V_0'|$ (here $V_0'\equiv dV_0(x)/dx$ with $x$ being the radial coordinate), exceeds the growth rate, $\gamma_{inst}$, in the absence of velocity shear. This corresponds to the condition that the Richardson number, $R_i\equiv(\gamma_{inst}/V_0')^2$, falls below unity, which is often used as an empirical rule for estimates of the shear flow effects in various experimental conditions. In this letter, we show, however, that the situation is more complex and, in general, the impact of velocity shear on fluid and plasma instabilities cannot be described just by the Richardson number. One such example  was given in Ref.~\onlinecite{zhanginfluence}. Here, we demonstrate that the responses of the instabilities to the external shear flow depend on the underlying mechanism of the instabilities. The Rayleigh-Taylor type instabilities, such as Rayleigh-Taylor (RT) instability in neutral fluids and the interchange modes (IM) in confined plasmas, are essentially two-dimensional aperiodic modes driven by the effective gravity. Therefore, the density perturbations in RT and charge separation in IM are directly affected by the external shear flow. In the resistive drift waves (RDW), which are dissipative negative energy modes, the electrostatic potential perturbations are largely maintained by the electron dynamics along the magnetic fields. As a result, the latter modes are much less affected by the velocity shear. 

One of the complications of the analysis of the velocity shear effect on fluid/plasma instabilities is the non-Hermitian nature of corresponding differential equations. As a result, the standard approach, where perturbed quantities, $\tilde{a}(\mathbf{r},t)$, are decomposed as $\tilde{a}(\mathbf{r},t)=\hat{a}(x)exp(-i\omega t+ik_yy+ik_zz)$, does not give the full set of the solutions of the linearized problem (e.g. see Refs. \onlinecite{tatsuno2001transient,trefethen2005spectra,MikhailenkoPRE2005} and the references therein). The other solutions belonging to the continuous spectrum are possible, which may grow non-exponentially in time.  Such the so-called non-modal solutions were investigated in a number of papers \cite{kelvin1887stability,hassam1992nonlinear,tatsuno2001transient,mikhailenko2002rayleigh,mikhailenko2000temporal} for both Rayleigh-Taylor type and drift waves instabilities assuming that the density gradients and flow shear, $V_0'$,  are constants. Then,  the initial value problem for  $\tilde{a}(\mathbf{r},t)$ was solved  in terms of the ``shearing box" variables $t,~\eta=y-V_0'xt,~x$ and $z$. In the linear approximation, such solutions may demonstrate transient time power-law amplifications of initial perturbations of $\tilde{a}(\mathbf{r},t)$ and may be important in situations when the modal (exponentially growing) instabilities are absent. 

In what follows we will only look for the solution of the linearized equations in the standard modal form. When exist, in the linear regime the solutions with  $\gamma=Im(\omega)>0$,  will dominate the transient non-modal solutions and, therefore, can be considered as the most important solutions. The roles of both modal and non-modal solutions in the nonlinear regime have to be analyzed separately and are not considered in our paper. Also, in order to avoid possible effects of the Kelvin-Helmholtz instability, we take $V_0(x)=V_0'x$ where $V_0'$ is constant. Moreover, we consider spatially  localized perturbations in plasma (or neutral  fluid) characterized  by the logarithmic gradient of equilibrium fluid/plasma density, $\Lambda_n(x)=-d\textup{ln}n_0/dx$. We will use the Boussinesq approximation and  we adopt the following $ansatz$ for the equilibrium density profile:
\begin{equation}
    \Lambda_n(x)= \frac{1}{2\Delta}\frac{\delta n}{\bar{n}}\frac{1}{cosh^2(x/\Delta)}\equiv \frac{1}{L_n}\frac{1}{cosh^2(x/\Delta)},\label{eq_density_profile}
\end{equation}
which for $\delta n <\bar{n}$ corresponds to $n_0(x)=\bar{n}-(\delta n/2)tanh(x/\Delta)$. 

We start with the Rayleigh-Taylor type (RT and IM) instabilities, where we will assume that the effective gravity acceleration $g$ is in the $x$-direction and take $k_z=0$. Then for perturbed stream function/electrostatic potential, $\tilde{\phi}=\phi(x)exp(-i\omega t+ik_yy)$, we have the following differential equation:
\begin{equation}
\frac{d^2 \phi}{d x^2}-k_y^2\phi-\Lambda_n(x)\left(\frac{gk_y^2}{\tilde{\omega}^2}+\sigma\frac{k_yV_0'}{\tilde{\omega}}\right)\phi=0,\label{eq-disper-interchange}
\end{equation}
where $\tilde{\omega}=\omega-k_yV_0'x$, and $\sigma=0$ for RT (e.g. see Ref.~\onlinecite{zhanginfluence}) while $\sigma=1$ for IM.

In the absence of velocity shear, Eq.~(\ref{eq-disper-interchange}) can be solved analytically by using similarity to the Schr\"odinger equation~\cite{landau2013quantum} for electron in the potential well $\propto -cosh^{-2}(x/\Delta)$, and we could find the growth rate, $\gamma$, versus the integer mode number, $m$. The fastest growing mode corresponds to $m=0$, where the eigenvalue is given by
\begin{equation}
-\omega^2=\gamma_{0}^2(\kappa)\equiv\bar{\gamma}^2|\kappa|/(1+|\kappa|),\label{eq-growth-interchange}    
\end{equation}
with $\kappa=\Delta\times k_y$, $\bar{\gamma}^2=g/L_n$, and the corresponding eigenfunction is $\phi_{0}=cosh^{-|\kappa|}(x/\Delta)$.

For small velocity shear, Eq.~(\ref{eq-disper-interchange}) can be solved on the basis of perturbation expansion by rewriting both deviations of frequency $\delta\omega=\omega-i\gamma_{0}$ and eigenfunction $\delta \phi=\phi(x)-\phi_{0}(x)$ in powers of $V_0'$, i.e., $\delta \omega =\sum_j \delta \omega_{j}(V_0')^j$ and $\delta \phi =\sum_j \delta \phi_{j}(V_0')^j$. With no restrictions we assume that $\phi_{0}(x)$ is real and limit our analysis by the second order of $V_0'$. After some algebra, we obtain
\begin{eqnarray}
\delta \omega_1&=&-\frac{\sigma \Delta}{2L_n}\frac{V_0'\kappa}{\kappa^2+|\kappa|}\label{eq-deltaomega2},\nonumber\\
\frac{\delta \omega_2}{\gamma_0}\mid_{\kappa\rightarrow\infty}&\approx&-\frac{i}{8}\frac{V_0'^2}{\bar{\gamma}^2}\kappa^2,\\
\frac{\delta \omega_2}{\gamma_{0}}\mid_{\kappa\rightarrow 0}&\approx& -\frac{i}{8|\kappa|}\frac{V_0'^2}{\bar{\gamma}^2}\left(\frac{\sigma\Delta^2}{L_n^2}+\pi^2|\kappa|^2\right),\nonumber
\end{eqnarray}
and the first order correction to the eigenfunction, $\delta \phi_1$, is antisymmetric and purely imaginary, which causes tilting of the eddies of equipotential contour in that the real part of $\tilde{\phi}$, $Re(\tilde{\phi})=\phi_0(x)cos(k_yy)-\delta\phi_1(x)sin(k_yy)$, will be tilted towards positive (negative) $x,y$-region for $V_0'>0$ ($V_0'<0$). Such tilting effect becomes more prominent for larger $|\kappa V_0'|$.

For large $|\kappa|$, Eq.~(\ref{eq-deltaomega2}) indicates that both RT and IM will be ``suppressed" when $|V_0'|\tilde{>}2\sqrt{2}\bar{\gamma}/|\kappa|$. The reduction of growth rate is mainly due to the first term in the bracket of Eq.~(\ref{eq-disper-interchange}). However, the velocity shear never eliminates the instability completely \cite{benilov2002does}. Taking into account that the width of the eigenfunction estimated from $\phi_0(\xi)$ is $\textup{w}/\Delta\sim 1/\sqrt{|\kappa|}\ll1$, $\Lambda_n(x)$ can be taken as unity that the non-modal approach~\cite{mikhailenko2002rayleigh} can be applied, which shows that the instability will be quenched when $V_0'\geq \bar{\gamma}/\sqrt{2}$. Therefore, for large $|\kappa|$, the reduction of the growth rate is more remarkable for the modal solutions than that for the non-modal solutions. Note that the expansion method is valid when $|V_0'k_y|\textup{w}/\bar{\gamma}\sim|V_0'|\sqrt{|\kappa|}/\bar{\gamma}<1$ and $\delta \omega_{1,2}<\bar{\gamma}$. Therefore, for $|\kappa|\gg 1$, the corresponding Eq.~(\ref{eq-deltaomega2}) is valid in the regime $|V_0'|< \sqrt{8}\bar{\gamma}/|\kappa|<\bar{\gamma}/\sqrt{|\kappa|}$, where the growth rate can be largely reduced by the velocity shear as shown in Eq.~(\ref{eq-deltaomega2}).

However, for $|\kappa|\rightarrow 0$, Eq.~(\ref{eq-deltaomega2}) shows that  the velocity shear has a stronger impact on the growth rate of IM than on that of RT. We also notice that the real frequency due to the velocity shear appears only in IM but not in RT. As a matter of fact, the case $\kappa\rightarrow 0$ corresponds to the eigenfunction extending along x-coordinate on the distance much larger than the width of $\Lambda_n(x)$ such that the density profile can be considered as a step-function. For this case, by integrating Eq.~(\ref{eq-disper-interchange}), we find
\begin{equation}
   \frac{\omega}{\gamma_0}=-\frac{\sigma\kappa}{2|\kappa|}\frac{V_0'}{\gamma_0}\frac{\Delta}{L_n}\pm i\sqrt{1-\frac{\sigma}{4}\left(\frac{V_0'}{\gamma_0}\frac{\Delta}{L_n}\right)^2}. \label{eq_interchange_smallk}
\end{equation}
Eq.~(\ref{eq_interchange_smallk}) is valid for any $V_0'$ providing $\kappa\rightarrow 0$ and consistent with Eq.~(\ref{eq-deltaomega2}) for small $V_0'$.

\begin{figure}[bt]
\centering
\begin{minipage}{0.45\textwidth}
\includegraphics[width=1\textwidth]{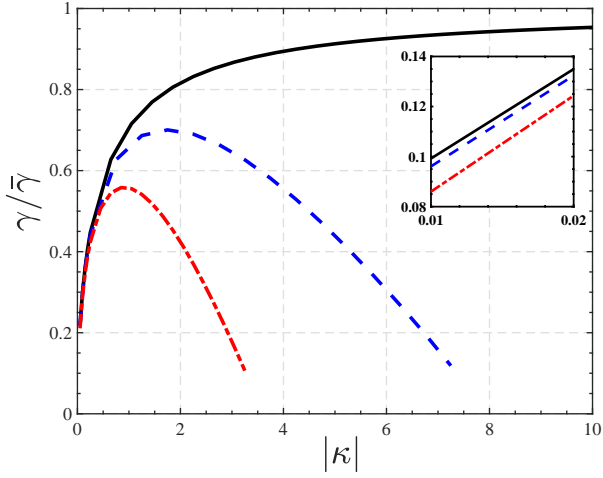}
\end{minipage}
\caption{The growth rate of the most unstable interchange mode versus $|\kappa|$ for $\Delta/L_n=\Delta n/2\bar{n}=1/4$ (inset is zoom-in for small $|\kappa|$). The curves of solid black, dashed blue, dash-dot red are for $V_0'/\bar{\gamma}=0,~0.2$ and $0.4$.}
\label{fig-interchange}
\end{figure}

\begin{figure}[hbt]
\centering
\begin{minipage}{0.45\textwidth}
\includegraphics[width=1\textwidth]{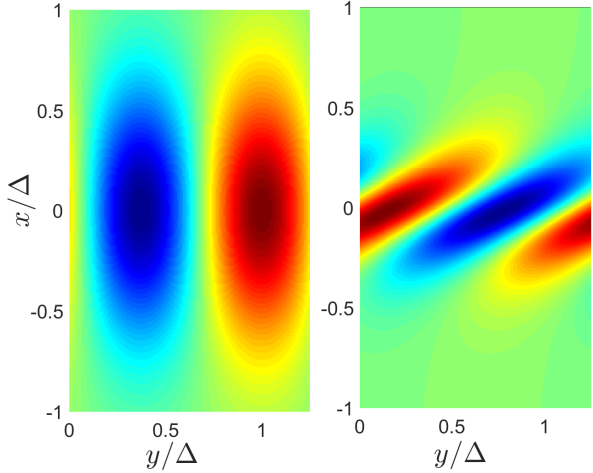}
\end{minipage}
\caption{Eddies of electrostatic potential contour for $\kappa=5$ and $\Delta/L_n=1/4$. $V_0'=0$ for the left panel and $V_0'=0.2\bar{\gamma}$ for the right panel.}
\label{fig-eddy-interchange}
\end{figure}

Numerical simulations of Eq.~(\ref{eq-disper-interchange}) are performed for RT/IM, which agree with our analyses. Here we only show the results of IM for illustration and those of RT can be found in Ref.~\onlinecite{zhanginfluence}, where the main difference is for $|\kappa|\ll 1$ as shown in Eq.~(\ref{eq-deltaomega2}). Moreover, in the simulations, we find that the choices of $\Delta/L_n$, the form\cite{benilov2002does} of $\Lambda_n(x)$ (e.g., the hyperbolic profile of $n_0(x)$ like in the pedestal region of tokamak\cite{mahdavi2002high} and in the edge of linear device\cite{carter2006intermittent}), and Boussinesq approximation don't affect the conclusions. Fig.~\ref{fig-interchange} depicts the growth rates of the most unstable modes of IM versus $|\kappa|$ for $\Delta/L_n=\Delta n/2\bar{n}=1/4$. It confirms that at large $|\kappa|$, the growth rate is significantly reduced in the presence of velocity shear, while for small $|\kappa|$, the reduction of the growth rate agrees with Eq.~(\ref{eq_interchange_smallk}). The eddies of equipotential contour $Re(\tilde{\phi})$ for the cases without and with velocity shear are plotted in Fig.~\ref{fig-eddy-interchange} and clearly demonstrate the tilting of the eddies due to the velocity shear. The real parts of $\omega$ from simulations agree with Eq.~(\ref{eq-deltaomega2}), illustrating that the eddies begin to propagate along y-direction due to the impact of velocity shear. 
 
Whereas the Rayleigh-Taylor type instabilities can be significantly suppressed by the velocity shear satisfying $|V_0'|>\gamma_{inst}$ ($R_i<1$), it's not the case for RDW, where the governing equation for $\phi(x)$ is
\begin{equation}
\rho_s^2\frac{d^2 \phi}{d x^2}-\left[ 1+\rho_s^2k_y^2-\frac{\omega_*(x)}{\tilde{\omega}}+i\frac{\tilde{\omega}-\omega_*(x)}{\nu_\parallel}
\right]\phi=0.\label{eq-disper-drift-wave}
\end{equation}
Here, $\rho_s=cT_e/eB_0\Omega_i$, $\Omega_i=eB_0/m_ic$, $\omega_*(x)=k_y \rho_s^2\Omega_i\Lambda_n(x)$, and $\nu_\parallel=k_z^2T_e/m\nu_{ei}$ with $\nu_{ei}$ being electron-ion collision frequency. The last term drives the instability by introducing a phase shift between the perturbations of electrostatic potential and density, where we have used the adiabatic limit $\nu_\parallel \gg \hat{\omega}_*\equiv k_y\rho_s^2\Omega_i/L_n$ to obtain Eq.~(\ref{eq-disper-drift-wave}) for RDW (the hydrodynamic limit $\nu_\parallel \ll \hat{\omega}_*$ is not relevant to Eq.~(\ref{eq-disper-drift-wave}) and thus is beyond the scope of this letter).

The dependence of growth rate on the mode number $m$ is complicated when solving Eq. (\ref{eq-disper-drift-wave}) by using the similarity to the Schr\"odinger equation. However, the most unstable mode of interest can be $m=0$ mode (e.g., for $\rho_sk_y=0.5$ and $\Delta/\rho_s=30$), which can largely simplify the analysis. Therefore, in the following, we will focus on the $m=0$ mode to elucidate the effect of velocity shear on RDW.

In the adiabatic limit we can assume that the driving term has little impact on the eigenfunctions but simply drives the instability, which allows us first finding $\phi$ from Eq.~(\ref{eq-disper-drift-wave}) without the driving term and then computing the growth rate by multiplying Eq.~(\ref{eq-disper-drift-wave}) with the complex conjugate of $\phi$ and integrating it over $x$-space:
\begin{equation}
\gamma\int\frac{\omega_*}{\tilde{\omega}_r^2}|\phi|^2dx=\int \frac{\omega_*-\tilde{\omega}_r}{\nu_\parallel}|\phi|^2dx,\label{eq_integral_imag}
\end{equation}
where $\gamma \ll \tilde{\omega}_r=\omega_r-k_yV_0(x)$ has been assumed with $\omega_r=Re(\omega)$.

Without velocity shear, Eq.~(\ref{eq-disper-drift-wave}) is similar to Eq.~(\ref{eq-disper-interchange}) and we obtain
\begin{equation}
\omega_{r}^0=\frac{\hat{\omega}_*}{(1+\rho^2k_y^2)(1+\epsilon^{-1})},~\phi_0(x)=cosh^{-\epsilon}(x/\Delta),\label{eq-growth-eigenfunction-drift}    
\end{equation}
where $\epsilon=(1+\rho_s^2k_y^2)^{1/2}\Delta\rho_s^{-1}\gg 1$. As a result, the growth rate can be estimated from Eq.~(\ref{eq_integral_imag}) as 
\begin{equation}
    \gamma_0=\frac{\hat{\omega}_*^2}{\nu_\parallel}\frac{\rho_s^2k_y^2+(2\epsilon)^{-1}}{(1+\rho_s^2k_y^2)^3},\label{eq-growth-dw-noshear}
\end{equation}
which is consistent with the result for constant $\omega_*$ when $\rho_s^2k_y^2\gg (2\epsilon)^{-1}$ (keeping in mind that the effective $\rho_sk_x\sim 2\pi(\rho_s/\Delta)^{1/2}\ll 1$ so it can be ignored in the growth rate for constant $\omega_*$).

From Eq.~(\ref{eq-growth-eigenfunction-drift}) wee see that the eigenfunction has localization width $\sim \sqrt{\rho_s\Delta}$, within which $cosh(x/\Delta)\approx exp[(x/\Delta)^2/2]$. As a result, for small $V_0'$, we have
\begin{equation}
  \frac{\omega_*}{\tilde{\omega}_r}\approx\frac{\hat{\omega}_*exp(x_0^2/\Delta^2)}{\omega_r~ cosh^2[(x-x_0)/\Delta]},\label{eq-omega_star-tilde_omega}
\end{equation}
where $x_0=k_y V_0'\Delta^2/2\omega_r^0$. Substituting it into Eq.~(\ref{eq-disper-drift-wave}) we find
\begin{equation}
\omega_r\approx \omega_r^0[1+k_y^2 V_0'^2\Delta^2/(2\omega_r^0)^2],~\phi=\phi_0(x-x_0),\label{eq-solution-correction-DW}
\end{equation}
which shows that the velocity shear simply shifts the eigenfunction toward positive $V_0$, without changing its shape (to the order of $\hat{\omega}_*/\nu_\parallel$). The growth rate can be estimated from the integral in Eq.~(\ref{eq_integral_imag}) as
\begin{equation}
    \gamma\approx \gamma_0[1-k_y^2 V_0'^2\Delta^2/2(\omega_r^0)^2].\label{eq-growth-rate-correction-final}
\end{equation}
Therefore, from Eqs.~(\ref{eq-solution-correction-DW}, \ref{eq-growth-rate-correction-final}) we see that the growth rate (real frequency) will quadratically decrease (increase) with $V_0'$.

Eq.~(\ref{eq-growth-rate-correction-final}) indicates that RDW will be stabilized when $|V_0'|>|V_0'|_{stab}\equiv\sqrt{2}\omega_r^0(\Delta k_y)^{-1}$. However, the localized solutions are only possible for $|V_0'|$ below some threshold value $|V_0'|_{loc}$, which is smaller than $|V_0'|_{stab}$. This effect can be interpreted by investigating the property of the potential well $U(x)=-\omega_*/\tilde{\omega}_r$ assuming that $k_yV_0'x\ll \omega_r\approx \omega_r^0$. As a result, the threshold value of $V_0'$ to have localized solution approximately corresponds to the transition of $U$ from a potential well to that with only one barrier at $V_0(x)<0$, provided that the effective ``energy" $E=-(1+\rho_s^2k_y^2)$ is close to the bottom of potential well, $-\hat{\omega}_*/\omega_r^0$, for zeroth mode. After some algebra, we obtain $|V_0'|_{loc}\approx 0.66\omega_r^0(\Delta k_y)^{-1}<|V_0'|_{stab}$ (note that no localized solution is possible beyond $|V_0'|_{loc}$ but non-modal solutions exist\cite{mikhailenko2000temporal} if $\Lambda_n$ can be taken as a constant). It follows that $\gamma\approx \gamma_0(1-0.22 V_0'^2/|V_0'|_{loc}^2)$. Therefore the strongest impact of velocity shear on the growth rate for the localized solution is $\delta \gamma=\gamma_0-\gamma\approx  0.22\gamma_0\ll \gamma_0$. Recalling the expression of $\gamma_0$, we have $\gamma_0\ll |V_0'|_{loc}$ when $\nu_\parallel\gg \hat{\omega}_*\rho_s^2\Delta k_y^3/(1+\rho_s^2k_y^2)^2$, which suggests that RDW cannot be significantly suppressed by the velocity shear even though $ |V_0'|\gg \gamma_{inst}$ corresponding to $R_i\ll 1$.

We note that at $V_0'=|V_0'|_{loc}$, $x_0\approx 0.33 \Delta$ and $k_yV_0'x_0\approx 0.22 \omega_r^0\ll \omega_r^0$. Therefore the assumption of $k_yV_0'x\ll \omega_r$ used in derivation of Eq.~(\ref{eq-omega_star-tilde_omega}) and in estimate of $|V_0'|_{loc}$ is valid. As a result, $\tilde{\omega}_r\sim \omega_r^0\gg \gamma$ such that the omission of $\gamma$ in Eq.~(\ref{eq_integral_imag}) is reasonable. We also note that an impact of magnetic shear and thus RDW coupling to the ion sound waves has been ignored in Eq.~(\ref{eq-disper-drift-wave}). However, the emission of sound waves leading to the dissipation occurs only at the wings of the eigenfunction and thus becomes small when $\Delta L_n<L_s\rho_s$, where $L_s$ is the magnetic shear length.

\begin{figure}[bt]
\centering
\begin{minipage}{0.45\textwidth}
\includegraphics[width=1\textwidth]{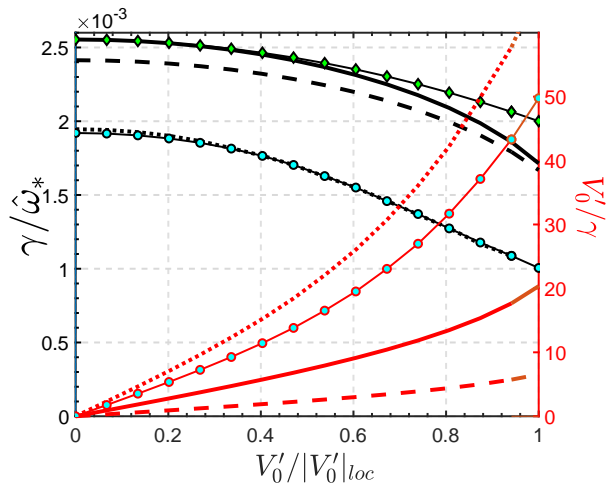}
\end{minipage}
\caption{The growth rates (black curves corresponding to the left y axis) and $V_0'/\gamma$ (red curves corresponding to the right y axis) of RDW versus $V_0'/|V_0'|_{loc}$ for $\nu_\parallel/\hat{\omega}_*=50$. The solid, doted and dashed curves are for $\rho_sk_y=0.5$, $\rho_sk_y=0.3$ and $\rho_sk_y=1$ with $\Delta=30\rho_s$. The curve with circle markers is for $\rho_sk_y=0.3$ but $\Delta=40\rho_s$. The curve with diamond markers is the fitting from Eq.~(\ref{eq-growth-rate-correction-final}) for the conditions of the solid curve.}
\label{fig-growth-drift-wave}
\end{figure}

\begin{figure}[hbt]
\centering
\begin{minipage}{0.45\textwidth}
\includegraphics[width=1\textwidth]{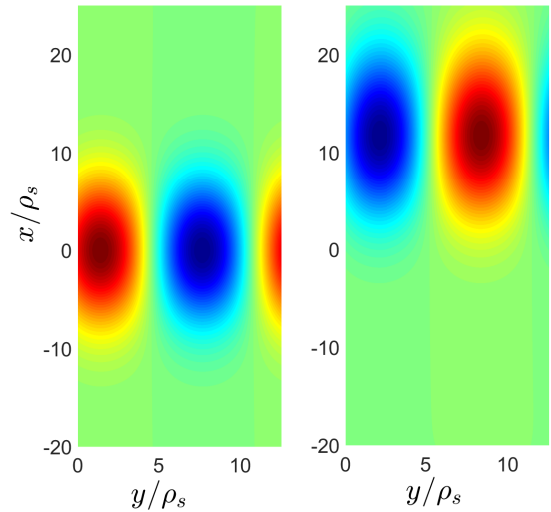}
\end{minipage}
\caption{Eddies of electrostatic potential contour corresponding to the eigenfunctions in Fig.~\ref{fig-growth-drift-wave} for $\rho_sk_y=0.5$ and $\Delta=30$. $V_0'=0$ for the left panel and $V_0'=0.8|V_0'|_{loc}$ for the right panel.}
\label{fig-eddy-drift-wave}
\end{figure}

Eq.~(\ref{eq-disper-drift-wave}) is solved numerically and we find that localized solutions are only possible for $|V_0'|\tilde{<}|V_0'|_{loc}$. The growth rate and ratio of $V_0'/\gamma$ of the most unstable mode are depicted as functions of $V_0'/|V_0'|_{loc}$ in Fig.~\ref{fig-growth-drift-wave} for $\nu_\parallel=50\hat{\omega}_*$ (the results are insensitive to $\nu_\parallel/\hat{\omega}_*\gg 1$). It confirms that the impact of velocity shear on the growth rate is small even though $|V_0'|\gg \gamma$, where the growth rates agrees well with Eq.~(\ref{eq-growth-rate-correction-final}) for all $V_0'<|V_0'|_{loc}$. We can also see that the growth rate as a function of $V_0'/|V_0'|_{loc}$ is strongly affected by $k_y$ but not by $\Delta$ as predicted by Eqs.~(\ref{eq-growth-dw-noshear},~\ref{eq-growth-rate-correction-final}). The eddies shown in Fig.~\ref{fig-eddy-drift-wave}, in agreement with our analytic results, are not tilted in the presence of velocity shear but simply shifted in the radial direction.

In conclusion, we investigated the influence of velocity shear on the localized RT/IM and RDW with inhomogeneous equilibrium fluid/plasma density. The effective ``potential well" induced by the inhomogeneous density allows us finding solutions in the localized domain. We find that the velocity shear with $|V_0'|>\gamma_{inst}$ has different effects on these instabilities: it can significantly suppress RT/IM but can only slightly reduce the growth rate of RDW. In addition, the velocity shear causes strong tilting of the eddies of the equipotential contour of RT/IM (see Fig.~\ref{fig-eddy-interchange}) but only shifts the eddies of RDW in the radial direction without tilting (see Fig.~\ref{fig-eddy-drift-wave}). However, for large velocity shear, the eigenmode solutions for RDW cease to exist.

These differences are due to the different physics of these instabilities. The RT instability is in regard to the dynamics of density variation and thus could be directly altered by the sheared flow. Similarly, IM modes are governed by the dynamics of plasma density perturbations with charge separation originated from the almost ``irreversible'' cross-field drift due to the gravity. Therefore, shear flow induced advection of density with embedded charges inevitably alters such instabilities. Whereas for RDW, the electric field largely results from the fast parallel electron dynamics in response to the plasma density perturbations in the adiabatic limit. Therefore, the distribution of electric charges virtually has no ``memory'' and, therefore, velocity shear makes a very small effect on the RDW instability.

The authors gratefully acknowledge fruitful discussions with P.H. Diamond. This work was supported by the U.S. Department of Energy, Office of Science, Office of Fusion Energy Sciences under Award No. DE-FG02-04ER54739 at UCSD.

\bibliography{main}% Produces the bibliography via BibTeX.

%merlin.mbs aipnum4-1.bst 2010-07-25 4.21a (PWD, AO, DPC) hacked
%Control: key (0)
%Control: author (8) initials jnrlst
%Control: editor formatted (1) identically to author
%Control: production of article title (0) allowed
%Control: page (1) range
%Control: year (1) truncated
%Control: production of eprint (0) enabled
\begin{thebibliography}{28}%
\makeatletter
\providecommand \@ifxundefined [1]{%
 \@ifx{#1\undefined}
}%
\providecommand \@ifnum [1]{%
 \ifnum #1\expandafter \@firstoftwo
 \else \expandafter \@secondoftwo
 \fi
}%
\providecommand \@ifx [1]{%
 \ifx #1\expandafter \@firstoftwo
 \else \expandafter \@secondoftwo
 \fi
}%
\providecommand \natexlab [1]{#1}%
\providecommand \enquote  [1]{``#1''}%
\providecommand \bibnamefont  [1]{#1}%
\providecommand \bibfnamefont [1]{#1}%
\providecommand \citenamefont [1]{#1}%
\providecommand \href@noop [0]{\@secondoftwo}%
\providecommand \href [0]{\begingroup \@sanitize@url \@href}%
\providecommand \@href[1]{\@@startlink{#1}\@@href}%
\providecommand \@@href[1]{\endgroup#1\@@endlink}%
\providecommand \@sanitize@url [0]{\catcode `\\12\catcode `\$12\catcode
  `\&12\catcode `\#12\catcode `\^12\catcode `\_12\catcode `\%12\relax}%
\providecommand \@@startlink[1]{}%
\providecommand \@@endlink[0]{}%
\providecommand \url  [0]{\begingroup\@sanitize@url \@url }%
\providecommand \@url [1]{\endgroup\@href {#1}{\urlprefix }}%
\providecommand \urlprefix  [0]{URL }%
\providecommand \Eprint [0]{\href }%
\providecommand \doibase [0]{http://dx.doi.org/}%
\providecommand \selectlanguage [0]{\@gobble}%
\providecommand \bibinfo  [0]{\@secondoftwo}%
\providecommand \bibfield  [0]{\@secondoftwo}%
\providecommand \translation [1]{[#1]}%
\providecommand \BibitemOpen [0]{}%
\providecommand \bibitemStop [0]{}%
\providecommand \bibitemNoStop [0]{.\EOS\space}%
\providecommand \EOS [0]{\spacefactor3000\relax}%
\providecommand \BibitemShut  [1]{\csname bibitem#1\endcsname}%
\let\auto@bib@innerbib\@empty
%</preamble>
\bibitem [{\citenamefont {Kelvin}(1887)}]{kelvin1887stability}%
  \BibitemOpen
  \bibfield  {author} {\bibinfo {author} {\bibfnamefont {L.}~\bibnamefont
  {Kelvin}},\ }\bibfield  {title} {\enquote {\bibinfo {title} {Stability of
  fluid motion: rectilinear motion of viscous fluid between two parallel
  plates},}\ }\href@noop {} {\bibfield  {journal} {\bibinfo  {journal} {Phil.
  Mag}\ }\textbf {\bibinfo {volume} {24}},\ \bibinfo {pages} {188--196}
  (\bibinfo {year} {1887})}\BibitemShut {NoStop}%
\bibitem [{\citenamefont {Kuo}(1963)}]{kuo1963perturbations}%
  \BibitemOpen
  \bibfield  {author} {\bibinfo {author} {\bibfnamefont {H.}~\bibnamefont
  {Kuo}},\ }\bibfield  {title} {\enquote {\bibinfo {title} {Perturbations of
  plane couette flow in stratified fluid and origin of cloud streets},}\
  }\href@noop {} {\bibfield  {journal} {\bibinfo  {journal} {The Physics of
  Fluids}\ }\textbf {\bibinfo {volume} {6}},\ \bibinfo {pages} {195--211}
  (\bibinfo {year} {1963})}\BibitemShut {NoStop}%
\bibitem [{\citenamefont
  {Chandrasekhar}(2013)}]{chandrasekhar2013hydrodynamic}%
  \BibitemOpen
  \bibfield  {author} {\bibinfo {author} {\bibfnamefont {S.}~\bibnamefont
  {Chandrasekhar}},\ }\href@noop {} {\emph {\bibinfo {title} {Hydrodynamic and
  hydromagnetic stability}}}\ (\bibinfo  {publisher} {Courier Corporation},\
  \bibinfo {year} {2013})\BibitemShut {NoStop}%
\bibitem [{\citenamefont {Hassam}(1992)}]{hassam1992nonlinear}%
  \BibitemOpen
  \bibfield  {author} {\bibinfo {author} {\bibfnamefont {A.}~\bibnamefont
  {Hassam}},\ }\bibfield  {title} {\enquote {\bibinfo {title} {Nonlinear
  stabilization of the {R}ayleigh-{T}aylor instability by external velocity
  shear},}\ }\href@noop {} {\bibfield  {journal} {\bibinfo  {journal} {Physics
  of Fluids B: Plasma Physics}\ }\textbf {\bibinfo {volume} {4}},\ \bibinfo
  {pages} {485--487} (\bibinfo {year} {1992})}\BibitemShut {NoStop}%
\bibitem [{\citenamefont {Benilov}, \citenamefont {Naulin},\ and\ \citenamefont
  {Rasmussen}(2002)}]{benilov2002does}%
  \BibitemOpen
  \bibfield  {author} {\bibinfo {author} {\bibfnamefont {E.}~\bibnamefont
  {Benilov}}, \bibinfo {author} {\bibfnamefont {V.}~\bibnamefont {Naulin}}, \
  and\ \bibinfo {author} {\bibfnamefont {J.~J.}\ \bibnamefont {Rasmussen}},\
  }\bibfield  {title} {\enquote {\bibinfo {title} {Does a sheared flow
  stabilize inversely stratified fluid?}}\ }\href@noop {} {\bibfield  {journal}
  {\bibinfo  {journal} {Physics of Fluids}\ }\textbf {\bibinfo {volume} {14}},\
  \bibinfo {pages} {1674--1680} (\bibinfo {year} {2002})}\BibitemShut {NoStop}%
\bibitem [{\citenamefont {Lehnert}(1966)}]{lehnert1966short}%
  \BibitemOpen
  \bibfield  {author} {\bibinfo {author} {\bibfnamefont {B.}~\bibnamefont
  {Lehnert}},\ }\bibfield  {title} {\enquote {\bibinfo {title} {Short-circuit
  of flute disturbances at a plasma boundary},}\ }\href@noop {} {\bibfield
  {journal} {\bibinfo  {journal} {The Physics of Fluids}\ }\textbf {\bibinfo
  {volume} {9}},\ \bibinfo {pages} {1367--1372} (\bibinfo {year}
  {1966})}\BibitemShut {NoStop}%
\bibitem [{\citenamefont {Sugama}\ and\ \citenamefont
  {Wakatani}(1991)}]{sugama1991radial}%
  \BibitemOpen
  \bibfield  {author} {\bibinfo {author} {\bibfnamefont {H.}~\bibnamefont
  {Sugama}}\ and\ \bibinfo {author} {\bibfnamefont {M.}~\bibnamefont
  {Wakatani}},\ }\bibfield  {title} {\enquote {\bibinfo {title} {Radial
  electric field effect on resistive interchange modes},}\ }\href@noop {}
  {\bibfield  {journal} {\bibinfo  {journal} {Physics of Fluids B: Plasma
  Physics}\ }\textbf {\bibinfo {volume} {3}},\ \bibinfo {pages} {1110--1112}
  (\bibinfo {year} {1991})}\BibitemShut {NoStop}%
\bibitem [{\citenamefont {Miller}\ \emph {et~al.}(1995)\citenamefont {Miller},
  \citenamefont {Waelbroeck}, \citenamefont {Hassam},\ and\ \citenamefont
  {Waltz}}]{miller1995stabilization}%
  \BibitemOpen
  \bibfield  {author} {\bibinfo {author} {\bibfnamefont {R.}~\bibnamefont
  {Miller}}, \bibinfo {author} {\bibfnamefont {F.}~\bibnamefont {Waelbroeck}},
  \bibinfo {author} {\bibfnamefont {A.}~\bibnamefont {Hassam}}, \ and\ \bibinfo
  {author} {\bibfnamefont {R.}~\bibnamefont {Waltz}},\ }\bibfield  {title}
  {\enquote {\bibinfo {title} {Stabilization of ballooning modes with sheared
  toroidal rotation},}\ }\href@noop {} {\bibfield  {journal} {\bibinfo
  {journal} {Physics of Plasmas}\ }\textbf {\bibinfo {volume} {2}},\ \bibinfo
  {pages} {3676--3684} (\bibinfo {year} {1995})}\BibitemShut {NoStop}%
\bibitem [{\citenamefont {Carreras}\ \emph {et~al.}(1993)\citenamefont
  {Carreras}, \citenamefont {Lynch}, \citenamefont {Garcia},\ and\
  \citenamefont {Diamond}}]{carreras1993resistive}%
  \BibitemOpen
  \bibfield  {author} {\bibinfo {author} {\bibfnamefont {B.}~\bibnamefont
  {Carreras}}, \bibinfo {author} {\bibfnamefont {V.}~\bibnamefont {Lynch}},
  \bibinfo {author} {\bibfnamefont {L.}~\bibnamefont {Garcia}}, \ and\ \bibinfo
  {author} {\bibfnamefont {P.}~\bibnamefont {Diamond}},\ }\bibfield  {title}
  {\enquote {\bibinfo {title} {Resistive pressure-gradient-driven turbulence
  with self-consistent flow profile evolution},}\ }\href@noop {} {\bibfield
  {journal} {\bibinfo  {journal} {Physics of Fluids B: Plasma Physics}\
  }\textbf {\bibinfo {volume} {5}},\ \bibinfo {pages} {1491--1505} (\bibinfo
  {year} {1993})}\BibitemShut {NoStop}%
\bibitem [{\citenamefont {Diamond}\ \emph {et~al.}(2005)\citenamefont
  {Diamond}, \citenamefont {Itoh}, \citenamefont {Itoh},\ and\ \citenamefont
  {Hahm}}]{diamond2005zonal}%
  \BibitemOpen
  \bibfield  {author} {\bibinfo {author} {\bibfnamefont {P.~H.}\ \bibnamefont
  {Diamond}}, \bibinfo {author} {\bibfnamefont {S.}~\bibnamefont {Itoh}},
  \bibinfo {author} {\bibfnamefont {K.}~\bibnamefont {Itoh}}, \ and\ \bibinfo
  {author} {\bibfnamefont {T.}~\bibnamefont {Hahm}},\ }\bibfield  {title}
  {\enquote {\bibinfo {title} {Zonal flows in plasma - a review},}\ }\href@noop
  {} {\bibfield  {journal} {\bibinfo  {journal} {Plasma Physics and Controlled
  Fusion}\ }\textbf {\bibinfo {volume} {47}},\ \bibinfo {pages} {R35} (\bibinfo
  {year} {2005})}\BibitemShut {NoStop}%
\bibitem [{\citenamefont {Waelbroeck}\ \emph {et~al.}(1992)\citenamefont
  {Waelbroeck}, \citenamefont {Antonsen~Jr}, \citenamefont {Guzdar},\ and\
  \citenamefont {Hassam}}]{waelbroeck1992theory}%
  \BibitemOpen
  \bibfield  {author} {\bibinfo {author} {\bibfnamefont {F.}~\bibnamefont
  {Waelbroeck}}, \bibinfo {author} {\bibfnamefont {T.}~\bibnamefont
  {Antonsen~Jr}}, \bibinfo {author} {\bibfnamefont {P.}~\bibnamefont {Guzdar}},
  \ and\ \bibinfo {author} {\bibfnamefont {A.}~\bibnamefont {Hassam}},\
  }\bibfield  {title} {\enquote {\bibinfo {title} {Theory of drift-acoustic
  instabilities in the presence of sheared flows},}\ }\href@noop {} {\bibfield
  {journal} {\bibinfo  {journal} {Physics of Fluids B: Plasma Physics}\
  }\textbf {\bibinfo {volume} {4}},\ \bibinfo {pages} {2441--2447} (\bibinfo
  {year} {1992})}\BibitemShut {NoStop}%
\bibitem [{\citenamefont {Ritz}\ \emph {et~al.}(1990)\citenamefont {Ritz},
  \citenamefont {Lin}, \citenamefont {Rhodes},\ and\ \citenamefont
  {Wootton}}]{ritz1990evidence}%
  \BibitemOpen
  \bibfield  {author} {\bibinfo {author} {\bibfnamefont {C.~P.}\ \bibnamefont
  {Ritz}}, \bibinfo {author} {\bibfnamefont {H.}~\bibnamefont {Lin}}, \bibinfo
  {author} {\bibfnamefont {T.}~\bibnamefont {Rhodes}}, \ and\ \bibinfo {author}
  {\bibfnamefont {A.~J.}\ \bibnamefont {Wootton}},\ }\bibfield  {title}
  {\enquote {\bibinfo {title} {Evidence for confinement improvement by
  velocity-shear suppression of edge turbulence},}\ }\href@noop {} {\bibfield
  {journal} {\bibinfo  {journal} {Physical Review Letters}\ }\textbf {\bibinfo
  {volume} {65}},\ \bibinfo {pages} {2543} (\bibinfo {year}
  {1990})}\BibitemShut {NoStop}%
\bibitem [{\citenamefont {Groebner}, \citenamefont {Burrell},\ and\
  \citenamefont {Seraydarian}(1990)}]{groebner1990role}%
  \BibitemOpen
  \bibfield  {author} {\bibinfo {author} {\bibfnamefont {R.}~\bibnamefont
  {Groebner}}, \bibinfo {author} {\bibfnamefont {K.}~\bibnamefont {Burrell}}, \
  and\ \bibinfo {author} {\bibfnamefont {R.}~\bibnamefont {Seraydarian}},\
  }\bibfield  {title} {\enquote {\bibinfo {title} {Role of edge electric field
  and poloidal rotation in the {L}-{H} transition},}\ }\href@noop {} {\bibfield
   {journal} {\bibinfo  {journal} {Physical Review Letters}\ }\textbf {\bibinfo
  {volume} {64}},\ \bibinfo {pages} {3015} (\bibinfo {year}
  {1990})}\BibitemShut {NoStop}%
\bibitem [{\citenamefont {Burrell}(1997)}]{burrell1997effects}%
  \BibitemOpen
  \bibfield  {author} {\bibinfo {author} {\bibfnamefont {K.}~\bibnamefont
  {Burrell}},\ }\bibfield  {title} {\enquote {\bibinfo {title} {Effects of
  {E}$\times${B} velocity shear and magnetic shear on turbulence and transport
  in magnetic confinement devices},}\ }\href@noop {} {\bibfield  {journal}
  {\bibinfo  {journal} {Physics of Plasmas}\ }\textbf {\bibinfo {volume} {4}},\
  \bibinfo {pages} {1499--1518} (\bibinfo {year} {1997})}\BibitemShut {NoStop}%
\bibitem [{\citenamefont {Van~Oost}\ \emph {et~al.}(2003)\citenamefont
  {Van~Oost}, \citenamefont {Adamek}, \citenamefont {Antoni}, \citenamefont
  {Balan}, \citenamefont {Boedo}, \citenamefont {Devynck}, \citenamefont
  {{\v{D}}uran}, \citenamefont {Eliseev}, \citenamefont {Gunn}, \citenamefont
  {Hron} \emph {et~al.}}]{van2003turbulent}%
  \BibitemOpen
  \bibfield  {author} {\bibinfo {author} {\bibfnamefont {G.}~\bibnamefont
  {Van~Oost}}, \bibinfo {author} {\bibfnamefont {J.}~\bibnamefont {Adamek}},
  \bibinfo {author} {\bibfnamefont {V.}~\bibnamefont {Antoni}}, \bibinfo
  {author} {\bibfnamefont {P.}~\bibnamefont {Balan}}, \bibinfo {author}
  {\bibfnamefont {J.}~\bibnamefont {Boedo}}, \bibinfo {author} {\bibfnamefont
  {P.}~\bibnamefont {Devynck}}, \bibinfo {author} {\bibfnamefont
  {I.}~\bibnamefont {{\v{D}}uran}}, \bibinfo {author} {\bibfnamefont
  {L.}~\bibnamefont {Eliseev}}, \bibinfo {author} {\bibfnamefont
  {J.}~\bibnamefont {Gunn}}, \bibinfo {author} {\bibfnamefont {M.}~\bibnamefont
  {Hron}},  \emph {et~al.},\ }\bibfield  {title} {\enquote {\bibinfo {title}
  {Turbulent transport reduction by {E}$\times${B} velocity shear during edge
  plasma biasing: recent experimental results},}\ }\href@noop {} {\bibfield
  {journal} {\bibinfo  {journal} {Plasma Physics and Controlled Fusion}\
  }\textbf {\bibinfo {volume} {45}},\ \bibinfo {pages} {621} (\bibinfo {year}
  {2003})}\BibitemShut {NoStop}%
\bibitem [{\citenamefont {Fujisawa}(2011)}]{fujisawa2011experimental}%
  \BibitemOpen
  \bibfield  {author} {\bibinfo {author} {\bibfnamefont {A.}~\bibnamefont
  {Fujisawa}},\ }\bibfield  {title} {\enquote {\bibinfo {title} {Experimental
  studies of mesoscale structure and its interactions with microscale waves in
  plasma turbulence},}\ }\href@noop {} {\bibfield  {journal} {\bibinfo
  {journal} {Plasma Physics and Controlled Fusion}\ }\textbf {\bibinfo {volume}
  {53}},\ \bibinfo {pages} {124015} (\bibinfo {year} {2011})}\BibitemShut
  {NoStop}%
\bibitem [{\citenamefont {Horton}(2018)}]{horton2012turbulent}%
  \BibitemOpen
  \bibfield  {author} {\bibinfo {author} {\bibfnamefont {W.}~\bibnamefont
  {Horton}},\ }\href@noop {} {\emph {\bibinfo {title} {Turbulent transport in
  magnetized plasmas}}}\ (\bibinfo  {publisher} {World Scientific, second
  edition},\ \bibinfo {year} {2018})\BibitemShut {NoStop}%
\bibitem [{\citenamefont {Kinsey}, \citenamefont {Waltz},\ and\ \citenamefont
  {Candy}(2005)}]{kinsey2005nonlinear}%
  \BibitemOpen
  \bibfield  {author} {\bibinfo {author} {\bibfnamefont {J.}~\bibnamefont
  {Kinsey}}, \bibinfo {author} {\bibfnamefont {R.}~\bibnamefont {Waltz}}, \
  and\ \bibinfo {author} {\bibfnamefont {J.}~\bibnamefont {Candy}},\ }\bibfield
   {title} {\enquote {\bibinfo {title} {Nonlinear gyrokinetic turbulence
  simulations of {E}$\times${B} shear quenching of transport},}\ }\href@noop {}
  {\bibfield  {journal} {\bibinfo  {journal} {Physics of Plasmas}\ }\textbf
  {\bibinfo {volume} {12}},\ \bibinfo {pages} {062302} (\bibinfo {year}
  {2005})}\BibitemShut {NoStop}%
\bibitem [{\citenamefont {Waltz}\ \emph {et~al.}(1997)\citenamefont {Waltz},
  \citenamefont {Staebler}, \citenamefont {Dorland}, \citenamefont {Hammett},
  \citenamefont {Kotschenreuther},\ and\ \citenamefont
  {Konings}}]{waltz1997gyro}%
  \BibitemOpen
  \bibfield  {author} {\bibinfo {author} {\bibfnamefont {R.}~\bibnamefont
  {Waltz}}, \bibinfo {author} {\bibfnamefont {G.}~\bibnamefont {Staebler}},
  \bibinfo {author} {\bibfnamefont {W.}~\bibnamefont {Dorland}}, \bibinfo
  {author} {\bibfnamefont {G.}~\bibnamefont {Hammett}}, \bibinfo {author}
  {\bibfnamefont {M.}~\bibnamefont {Kotschenreuther}}, \ and\ \bibinfo {author}
  {\bibfnamefont {J.}~\bibnamefont {Konings}},\ }\bibfield  {title} {\enquote
  {\bibinfo {title} {A gyro-{L}andau-fluid transport model},}\ }\href@noop {}
  {\bibfield  {journal} {\bibinfo  {journal} {Physics of Plasmas}\ }\textbf
  {\bibinfo {volume} {4}},\ \bibinfo {pages} {2482--2496} (\bibinfo {year}
  {1997})}\BibitemShut {NoStop}%
\bibitem [{\citenamefont {Zhang}, \citenamefont {Krasheninnikov},\ and\
  \citenamefont {Smolyakov}()}]{zhanginfluence}%
  \BibitemOpen
  \bibfield  {author} {\bibinfo {author} {\bibfnamefont {Y.}~\bibnamefont
  {Zhang}}, \bibinfo {author} {\bibfnamefont {S.~I.}\ \bibnamefont
  {Krasheninnikov}}, \ and\ \bibinfo {author} {\bibfnamefont {A.~I.}\
  \bibnamefont {Smolyakov}},\ }\bibfield  {title} {\enquote {\bibinfo {title}
  {Influence of flow shear on localized {R}ayleigh--{T}aylor and resistive
  drift wave instabilities},}\ }\href@noop {} {\bibinfo  {journal}
  {Contributions to Plasma Physics. 2019; e201900098}\ }\BibitemShut {NoStop}%
\bibitem [{\citenamefont {Tatsuno}, \citenamefont {Volponi},\ and\
  \citenamefont {Yoshida}(2001)}]{tatsuno2001transient}%
  \BibitemOpen
\bibfield  {journal} {  }\bibfield  {author} {\bibinfo {author} {\bibfnamefont
  {T.}~\bibnamefont {Tatsuno}}, \bibinfo {author} {\bibfnamefont
  {F.}~\bibnamefont {Volponi}}, \ and\ \bibinfo {author} {\bibfnamefont
  {Z.}~\bibnamefont {Yoshida}},\ }\bibfield  {title} {\enquote {\bibinfo
  {title} {Transient phenomena and secularity of linear interchange
  instabilities with shear flows in homogeneous magnetic field plasmas},}\
  }\href@noop {} {\bibfield  {journal} {\bibinfo  {journal} {Physics of
  Plasmas}\ }\textbf {\bibinfo {volume} {8}},\ \bibinfo {pages} {399--406}
  (\bibinfo {year} {2001})}\BibitemShut {NoStop}%
\bibitem [{\citenamefont {Trefethen}\ and\ \citenamefont
  {Embree}(2005)}]{trefethen2005spectra}%
  \BibitemOpen
  \bibfield  {author} {\bibinfo {author} {\bibfnamefont {L.~N.}\ \bibnamefont
  {Trefethen}}\ and\ \bibinfo {author} {\bibfnamefont {M.}~\bibnamefont
  {Embree}},\ }\href@noop {} {\emph {\bibinfo {title} {Spectra and
  pseudospectra: the behavior of nonnormal matrices and operators}}}\ (\bibinfo
   {publisher} {Princeton University Press},\ \bibinfo {year}
  {2005})\BibitemShut {NoStop}%
\bibitem [{\citenamefont {Mikhailenko}, \citenamefont {Scime},\ and\
  \citenamefont {Mikhailenko}(2005)}]{MikhailenkoPRE2005}%
  \BibitemOpen
  \bibfield  {author} {\bibinfo {author} {\bibfnamefont {V.~S.}\ \bibnamefont
  {Mikhailenko}}, \bibinfo {author} {\bibfnamefont {E.~E.}\ \bibnamefont
  {Scime}}, \ and\ \bibinfo {author} {\bibfnamefont {V.~V.}\ \bibnamefont
  {Mikhailenko}},\ }\bibfield  {title} {\enquote {\bibinfo {title} {Stability
  of stratified flow with inhomogeneous shear},}\ }\href@noop {} {\bibfield
  {journal} {\bibinfo  {journal} {Physical Review E}\ }\textbf {\bibinfo
  {volume} {71}},\ \bibinfo {pages} {026306} (\bibinfo {year}
  {2005})}\BibitemShut {NoStop}%
\bibitem [{\citenamefont {Mikhailenko}, \citenamefont {Mikhailenko},\ and\
  \citenamefont {Weiland}(2002)}]{mikhailenko2002rayleigh}%
  \BibitemOpen
  \bibfield  {author} {\bibinfo {author} {\bibfnamefont {V.~S.}\ \bibnamefont
  {Mikhailenko}}, \bibinfo {author} {\bibfnamefont {V.~V.}\ \bibnamefont
  {Mikhailenko}}, \ and\ \bibinfo {author} {\bibfnamefont {J.}~\bibnamefont
  {Weiland}},\ }\bibfield  {title} {\enquote {\bibinfo {title}
  {Rayleigh--{T}aylor instability in plasmas with shear flow},}\ }\href@noop {}
  {\bibfield  {journal} {\bibinfo  {journal} {Physics of Plasmas}\ }\textbf
  {\bibinfo {volume} {9}},\ \bibinfo {pages} {2891--2895} (\bibinfo {year}
  {2002})}\BibitemShut {NoStop}%
\bibitem [{\citenamefont {Mikhailenko}, \citenamefont {Mikhailenko},\ and\
  \citenamefont {Stepanov}(2000)}]{mikhailenko2000temporal}%
  \BibitemOpen
  \bibfield  {author} {\bibinfo {author} {\bibfnamefont {V.}~\bibnamefont
  {Mikhailenko}}, \bibinfo {author} {\bibfnamefont {V.}~\bibnamefont
  {Mikhailenko}}, \ and\ \bibinfo {author} {\bibfnamefont {K.}~\bibnamefont
  {Stepanov}},\ }\bibfield  {title} {\enquote {\bibinfo {title} {Temporal
  evolution of linear drift waves in a collisional plasma with homogeneous
  shear flow},}\ }\href@noop {} {\bibfield  {journal} {\bibinfo  {journal}
  {Physics of Plasmas}\ }\textbf {\bibinfo {volume} {7}},\ \bibinfo {pages}
  {94--100} (\bibinfo {year} {2000})}\BibitemShut {NoStop}%
\bibitem [{\citenamefont {Landau}\ and\ \citenamefont
  {Lifshitz}(2013)}]{landau2013quantum}%
  \BibitemOpen
  \bibfield  {author} {\bibinfo {author} {\bibfnamefont {L.}~\bibnamefont
  {Landau}}\ and\ \bibinfo {author} {\bibfnamefont {E.}~\bibnamefont
  {Lifshitz}},\ }\href@noop {} {\emph {\bibinfo {title} {Quantum mechanics:
  non-relativistic theory}}},\ Vol.~\bibinfo {volume} {3}\ (\bibinfo
  {publisher} {Elsevier},\ \bibinfo {year} {2013})\BibitemShut {NoStop}%
\bibitem [{\citenamefont {Mahdavi}\ \emph {et~al.}(2002)\citenamefont
  {Mahdavi}, \citenamefont {Osborne}, \citenamefont {Leonard}, \citenamefont
  {Chu}, \citenamefont {Doyle}, \citenamefont {Fenstermacher}, \citenamefont
  {McKee}, \citenamefont {Staebler}, \citenamefont {Petrie}, \citenamefont
  {Wade} \emph {et~al.}}]{mahdavi2002high}%
  \BibitemOpen
  \bibfield  {author} {\bibinfo {author} {\bibfnamefont {M.}~\bibnamefont
  {Mahdavi}}, \bibinfo {author} {\bibfnamefont {T.}~\bibnamefont {Osborne}},
  \bibinfo {author} {\bibfnamefont {A.}~\bibnamefont {Leonard}}, \bibinfo
  {author} {\bibfnamefont {M.}~\bibnamefont {Chu}}, \bibinfo {author}
  {\bibfnamefont {E.}~\bibnamefont {Doyle}}, \bibinfo {author} {\bibfnamefont
  {M.}~\bibnamefont {Fenstermacher}}, \bibinfo {author} {\bibfnamefont
  {G.}~\bibnamefont {McKee}}, \bibinfo {author} {\bibfnamefont
  {G.}~\bibnamefont {Staebler}}, \bibinfo {author} {\bibfnamefont
  {T.}~\bibnamefont {Petrie}}, \bibinfo {author} {\bibfnamefont
  {M.}~\bibnamefont {Wade}},  \emph {et~al.},\ }\bibfield  {title} {\enquote
  {\bibinfo {title} {High performance {H} mode plasmas at densities above the
  greenwald limit},}\ }\href@noop {} {\bibfield  {journal} {\bibinfo  {journal}
  {Nuclear Fusion}\ }\textbf {\bibinfo {volume} {42}},\ \bibinfo {pages} {52}
  (\bibinfo {year} {2002})}\BibitemShut {NoStop}%
\bibitem [{\citenamefont {Carter}(2006)}]{carter2006intermittent}%
  \BibitemOpen
  \bibfield  {author} {\bibinfo {author} {\bibfnamefont {T.}~\bibnamefont
  {Carter}},\ }\bibfield  {title} {\enquote {\bibinfo {title} {Intermittent
  turbulence and turbulent structures in a linear magnetized plasma},}\
  }\href@noop {} {\bibfield  {journal} {\bibinfo  {journal} {Physics of
  plasmas}\ }\textbf {\bibinfo {volume} {13}},\ \bibinfo {pages} {010701}
  (\bibinfo {year} {2006})}\BibitemShut {NoStop}%
\end{thebibliography}%

\end{document}